\documentclass[journal]{IEEEtran}
\usepackage{setspace}
\usepackage{amssymb}
\usepackage{stmaryrd}
\usepackage{algorithmic}
\usepackage{tikz}
\usetikzlibrary{shapes}
\usepackage{tabularx}
\usepackage{multirow}
\usepackage{arydshln}
\usepackage{lscape}  
\usepackage[justification=raggedright]{caption}	
\usepackage{acro}
\acsetup{first-style=short}
\usepackage{array}
\newcolumntype{x}[1]{>{\centering\arraybackslash\hspace{0pt}}p{#1}}
\usepackage{color}

\newcounter{Q}
\newcounter{TD}

  \graphicspath{{Figures/}}
  \DeclareGraphicsExtensions{.pdf,.jpeg,.png}
\usepackage{graphicx}

\usepackage{epstopdf}
\usepackage{caption}
\usepackage{subcaption}
\usepackage{amsmath}

\usepackage{authblk}

\begin{document}
%
\title{Video-Capable Ultrasonic Wireless Communications through Biological Tissues}
%
%
%

\author[1,2]{Gizem Tabak}
\author[1,2]{Sijung Yang}
\author[2]{Rita Miller}
\author[2]{Michael Oelze}
\author[1,2]{Andrew Singer}
\affil[1]{Coordinated Science Laboratory, University of Illinois at Urbana Champaign Urbana, IL USA}
\affil[2]{Beckman Institute, University of Illinois at Urbana Champaign Urbana, IL USA}
\affil[ ]{E-mail:\{tabak2, syang103, rjmille, oelze, acsinger\}@illinois.edu}

\maketitle

\begin{abstract}
The use of wireless implanted medical devices (IMDs) is growing because they facilitate monitoring of patients at home and during normal activities, reduce the discomfort of patients and reduce the likelihood of infection associated with trailing wires. Currently, radiofrequency (RF) electromagnetic waves are the most commonly used method for communicating wirelessly with IMDs. However, due to the restrictions on the available bandwidth and the employable power, data rates of RF-based IMDs are limited to 267 kbps \cite{koprowski2015overview}. Considering standard definition video streaming requires data rates of 1.2 mbps and high definition requires 3 mbps, it is not possible to use the RF electromagnetic communications for high data rate communication applications such as video streaming. In this work, an alternative method that utilizes ultrasonic waves to relay information at high data rates is introduced. An advanced quadrature amplitude modulation (QAM) modem with phase-compensating, sparse decision feedback equalizer (DFE) is tailored to realize the full potential of the ultrasonic channel through biological tissues. The proposed system is tested on simulated experiments with finite impulse response (FIR) channel models and physical experiments with biological tissues. Consequently, the simulations and experiments demonstrated that video-capable data rates can be achieved with milimeter-sized transducers communicating through simulated FIR channels, through \textit{ex vivo} beef liver and \textit{ex vivo} pork chop samples, and through \textit{in situ} rabbit abdomen.
\end{abstract}

%
%
\IEEEpeerreviewmaketitle

\section{Introduction}
%
%
%
%

\IEEEPARstart{M}{any} modern wireless implanted medical devices (IMDs) make use of sensors within the body and communicate data wirelessly to a receiver or transmitter outside of the body. Because of technological advances, these devices are rapidly becoming an integral part of medical diagnostic and treatment procedures. About 1 in 10 people in America and about 1 in 17 people living in industrialized countries rely on IMDs to regain body function, to improve life quality or to maintain survival \cite{jiang2009technology}. Applications employing IMDs include, but are not limited to, pacemakers that prevent cardiovascular malfunctions, insulin monitors and pumps that control glucose levels in the blood and adjust insulin levels accordingly, and capsule endoscopy cameras that record the digestive tract when swallowed and deliver diagnostic information about gastrointestinal conditions. Some of these devices, such as pacemakers, are designed to perform a task to overcome deficiencies of the patient’s body and to be replaced invasively once their batteries are exhausted. Other devices, such as ingestible cameras, are designed to be collected after data acquisition, and the data can be processed offline. Nevertheless, these systems are not connected devices that relay high-bandwidth, real-time information. Therefore, they lack the capability of instantaneous, \textit{in situ} intervention. They have to be followed up with invasive, interventional procedures in case an anomaly is detected, and a delay in necessary clinical intervention can result in declining patient outcomes. Hence, a medically significant need exists to develop an active and wirelessly communicating system that can relay information in real-time or near real-time to devices outside of the body and open up the possibility of instantaneous intervention.

\begin{table*}[h]
\begin{minipage}{\textwidth}
\captionof{table}{Comparison of recent works employing ultrasound for communicating with IMDs}
\centering
\label{tab:literature}
\renewcommand{\arraystretch}{1.5}
\begin{tabular}{l|c|c|c|c|c|c|}
\cline{2-7}
                                                & \textbf{Transmission Medium} & \textbf{Transducer Size}   & \textbf{Modulation}   & \textbf{Equalizer}                                                                  & \textbf{Data Rates} & \textbf{BER}    \\ \hline
\multicolumn{1}{|l|}{\cite{bos2018enabling}}                       & Phantom+Scatter              & 2 mm                        & QPSK                  & -                                                                                   & 200 kbps            & \textless{}1e-4 \\ \hline
\multicolumn{1}{|l|}{\cite{wang2017exploiting}}                      & Mineral oil                  & 1.1x1.1x1.4 mm$^3$          & QPSK (2x2 link)       & -                                                                                   & 2x125 kbps          & \textless{}1e-4 \\ \hline
\multicolumn{1}{|l|}{\cite{chang201727}}                     & Castor oil\footnote{Experiments are also performed with animal tissue; but the corresponding data rates are not reported.}                  & 0.6x0.6x0.4 mm$^3$          & OOK                   & -                                                                                   & 95 kbps             & \textless{}1e-4 \\ \hline
\multicolumn{1}{|l|}{\multirow{2}{*}{\cite{kondapalli2017multiaccess}}}   & Chicken phantom              & 1 mm                        & OOK/CDMA              & -                                                                                   & 1 mbps              & 1e-02        \\ \cdashline{2-7} 
\multicolumn{1}{|l|}{}                          & Live ovine                   & 1 mm                        & OOK/CDMA              & -                                                                                   & 800 kbps            & 1e-01        \\ \hline
\multicolumn{1}{|l|}{\multirow{2}{*}{\cite{demirors2016high}}} & \multirow{2}{*}{Phantom}     & \multirow{2}{*}{19x44 mm$^2$} & \multirow{2}{*}{OFDM} & \multirow{2}{*}{\begin{tabular}[c]{@{}c@{}}Channel estimation\\ based\end{tabular}} & 28.12 mbps          & 1.3e-01        \\ \cdashline{6-7} 
\multicolumn{1}{|l|}{}                          &                              &                            &                       &                                                                                     & 12.09 mbps          & 1.9e-04        \\ \hline
\multicolumn{1}{|l|}{\cite{santagati2014sonar}}                 & Kidney phantom               & 9.5 mm                      & PPM                   & -                                                                                   & 70-700 kbps         & \textless{}1e-6 \\ \hline
\multicolumn{1}{|l|}{\multirow{2}{*}{\cite{singer2016mbps}}}   & Beef liver              &  19x39 mm$^2$                        &        64-QAM       & Adaptive equalizer                                                                                   & 20 mbps              & \textless{}1e-04        \\ \cdashline{2-7} 
\multicolumn{1}{|l|}{}                          & Pork chop                  &  19x39 mm$^2$                        &       64-QAM        & Adaptive equalizer                                                                                    &  30 mbps            & \textless{}1e-04        \\ \hline
\end{tabular}
\vspace{-.3cm}
\end{minipage}%
\end{table*}%

Currently, radio-frequency (RF) electromagnetic waves are the most frequently used method in wireless communication applications such as television, radio, or mobile phone communications. When RF waves travel through the air, they experience little attenuation. Additionally, they can operate at high frequencies, where the available bandwidth is also high. Their capability of operating at high frequencies while experiencing low loss makes RF waves appropriate for long-range, high data rate wireless communication applications through the air. However, there are various drawbacks of using RF waves with wireless IMDs to transmit data through the body. First, RF waves are highly attenuated in the body and have limited penetration depth. RF waves can travel 10 centimeters through the body from a deep-tissue IMD before experiencing 60 dB of path loss \cite{sayrafian2009statistical}. In stark contrast, they can travel as far as 59 meters through air before undergoing the same loss at the same frequency \cite{IEEEstandard}. Therefore, higher power levels need to be employed to compensate for the losses in the body due to high attenuation. However, the RF signal power levels that an IMD could deploy are limited for safety reasons, as higher power increases the risk of tissue damage \cite{FCCtelemetry}. There are also federal regulations on the allocation of the RF spectrum use within, and outside the body. The Medical Device Radio Communications (MedRadio) guidelines impose different rules on wireless implanted medical devices (IMD), medical body area networks (MBAN) and medical micro-power networks (MMN). While the first include the devices designed for communicating with an implant inside the body, MBAN include the network of sensors that are worn on the body, and MMN include implanted devices that help restore functions to limbs and organs. According to MedRadio guidelines, the allocated operation frequencies for the IMDs are within the range of 401-406 MHz, and the corresponding maximum allowed bandwidth is 300 kHz \cite{FCCradio}. Moreover, MedRadio transceivers are further limited by interference regulations because they must be able to operate in the presence of primary and secondary users in those bands. Such restrictions on the transmit power and operable bandwidth lead to fundamental performance constraints for IMDs employing RF links and the data rates of the current RF-based IMDs are demonstrated to be limited to 267 kbps \cite{koprowski2015overview}. Considering, for example, the standard definition video requires 1.2 Mbps bitrate, while high definition video streaming starts at 3 Mbps \cite{IBM}, these regulations set a significant barrier against possible wireless IMD applications to include video transmission. 

For many years, ultrasonic waves have been widely used as an alternative to RF electromagnetic waves for in underwater communication applications, where RF waves experience significant losses. Employing ultrasonic waves for such applications enables data rates of 1.2 mbps over 12 m under water \cite{riedl2014towards}, as opposed to 50 kbps over similar distances with electromagnetic waves \cite{palmeiro2011underwater}. Because acoustic waves have been used broadly in underwater communication applications for a long time, the characteristics of the underwater acoustic communication channel have been well established. The underwater acoustic communication channel is time-varying due to motion and the changing environment, dispersive due to speed and attenuation values that vary with frequency, temperature, salinity and pressure, frequency selective with long delay spread due to multipath, and with Doppler effects due to motion. Considering a wireless medical implant moving inside the body and communicating with a receiving probe outside of the body, most if not all of these characteristics apply to the through-body ultrasonic communication channel as well. Hence, employing advanced underwater communication techniques is a promising approach in order to achieve video-capable data rates through biological tissues. 

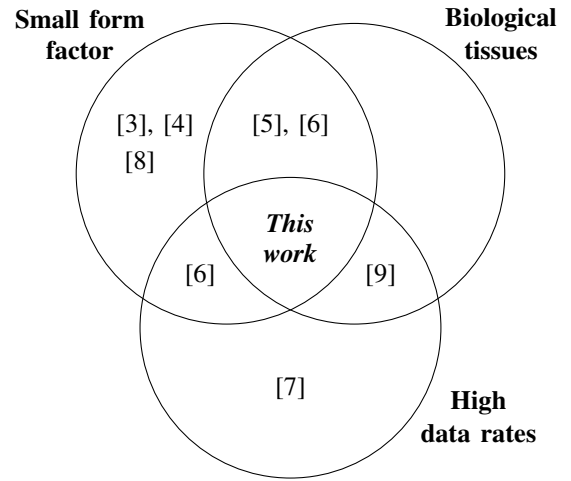
\begin{figure}[t]
\centering
\def\firstcircle{(135:1.2) circle (2)}
\def\secondcircle{(45:1.2) circle (2)}
\def\thirdcircle{(-90:1.2) circle (2)}
\begin{tikzpicture}
    \draw \firstcircle;
    \draw \secondcircle;
    \draw \thirdcircle;
    \node at (0,-2) {\cite{demirors2016high}};
    \node at (0,1.5) {\cite{chang201727}, \cite{kondapalli2017multiaccess}};
    \node at (1.2,-0.5) {\cite{singer2016mbps}};
    \node at (-1.2,-0.5) {\cite{kondapalli2017multiaccess}};
    \node at (-1.8,1.5) {\cite{bos2018enabling}, \cite{wang2017exploiting}};
    \node at (-2,1) {\cite{santagati2014sonar}};
    \node at (2.5,-2.2) {\textbf{High}};
    \node at (2.5,-2.6) {\textbf{data rates}};
    \node at (2.8,2.9) {\textbf{Biological}};
    \node at (2.8,2.5) {\textbf{tissues}};
    \node at (-2.8,2.9) {\textbf{Small form}};
    \node at (-2.8,2.5) {\textbf{factor}};
    \node at (0,0.2) {\textbf{\textit{This}}};
    \node at (0,-0.2) {\textbf{\textit{work}}};
\end{tikzpicture}
\caption{This is the first work in the literature that achieves high data rates ($>$1 mbps) through real biological tissues (as opposed to phantoms) using small form factor ($<$1 cm) transducers.}\label{fig:venn}
\end{figure}

The similarities between the ultrasonic communication channels through the body and under water suggest ultrasonic waves as a promising option for high data rate transmission \textit{in situ}. Besides experiencing lower loss and hence propagating deeper in the tissue, ultrasonic waves are desirable for wireless through-body communications for several other reasons: First, because the loss is lower compared to RF electromagnetic waves, the transmission can take place at lower transmit power levels. As a result, the patient experiences lower, if not insignificant, tissue heating. Second, medical applications that utilize ultrasonic waves, such as ultrasonic imaging, have been considered as a safer option compared to applications that utilize electromagnetic waves, such as X-ray imaging, which exposes the patients to significant amounts of ionizing radiation \cite{FDA}. Third, because there are no official regulations on the ultrasonic frequency spectrum, the available bandwidth, and the corresponding potential for high data rates, are significantly higher. For all these reasons, employing ultrasonic waves for through-tissue communications at video-capable data rates offers a safe and efficient alternative to RF communications.

\begin{figure*}[t]
\includegraphics[width=\textwidth]{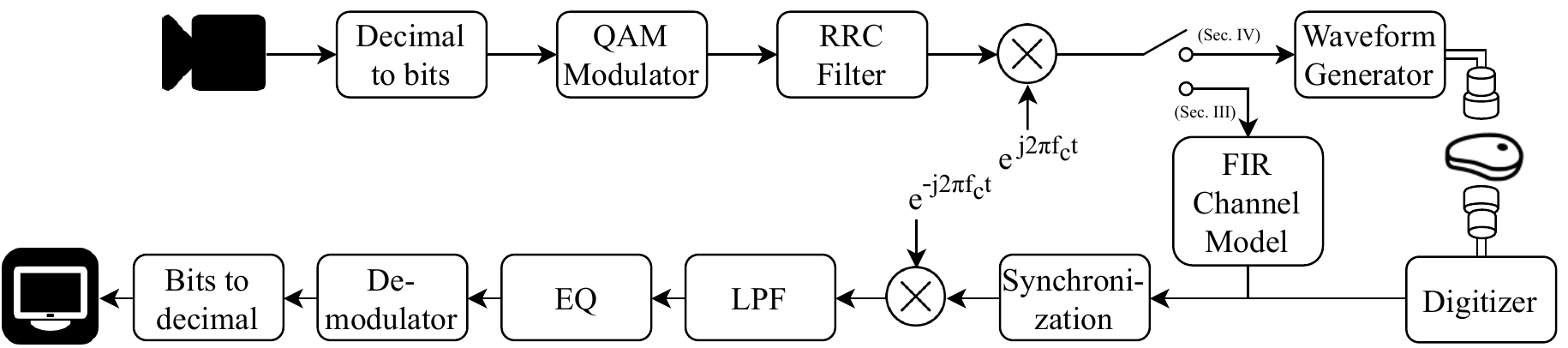}
\caption{Proposed communication system with equalizer (EQ) at the receiver}
\label{fig:framework}
\end{figure*}

Ultrasonic waves have been used in the literature for wireless in-body and through-body communications and have demonstrated the feasibility of the ultrasonic communication link through biological tissues. A comparison of recent works in the literature employing ultrasound for through-tissue communications is provided in Table \ref{tab:literature}. Nevertheless, the methods in the literature either achieve lower data rates ($<$1 mbps) that are insufficient for standard video communication \cite{bos2018enabling}, \cite{wang2017exploiting}, \cite{chang201727}, \cite{kondapalli2017multiaccess}, \cite{santagati2014sonar} or achieve higher data rates with large form factor transducers ($>$1cm) that could not be utilized in a small implantable device \cite{demirors2016high}, \cite{singer2016mbps}, or the communication link is established through phantoms instead of real biological tissues \cite{kondapalli2017multiaccess}, \cite{demirors2016high}. To the authors' knowledge, this work is the first work to demonstrate video-capable data rates with small form factor transducers through real \textit{ex vivo} and \textit{in situ} biological tissues (Fig. \ref{fig:venn}). The contributions of this work can be summarized as follows:
\begin{itemize}
    \item In order to utilize the ultrasonic through-tissue communication channel more efficiently, several underwater acoustic communication techniques are tailored to this particular application. At the transmitter end, a high order, spectrally efficient modulation technique (quadrature amplitude modulation (QAM)) is utilized to relay information at video-capable data rates. At the receiver end, a phase-tracking, sparse decision feedback equalizer (DFE) \cite{stojanovic1994phase}, \cite{lopez2001dfe} is used in order to compensate for the distortion and the intersymbol interference (ISI) introduced by the channel, and to recover the transmitted data successfully at high data rates.
    \item Simulated experiments are performed with finite impulse response (FIR) channel measurements provided in \cite{bos2018enabling} in order to examine the theoretical limits and to demonstrate the improvement provided by the proposed method over the previous methods in the literature. The simulations test the method with different channel types and demonstrate 3 to 10 fold increase in data rates with the same signal-to-noise ratios (SNR) and bit error rates (BER) compared to a basic QAM modem.
    \item An experimental test platform is set up with modular transducers through similar transmission media. For four different transducers, data is transmitted at different rates through water and through different thicknesses of \textit{ex vivo} beef liver slices. The experiments demonstrate the capabilities of the system with different transducers, which have different center frequencies and sizes, communicating through the biological tissue at video-capable rates.
    \item In order to investigate the capability of the proposed system through more complex communication channels, experiments are repeated with small, biocompatible transducers transmitting through \textit{ex vivo} pork chops and \textit{in situ} through the rabbit abdomen. Video-capable data rates are achieved with BER\textless1e-3 through these more complex transmission channels with uncoded transmission, and this BER can be made arbitrarily low using modest forward error correction (FEC)).
\end{itemize}

In an effort to elaborate on these contributions, the rest of the paper proceeds as follows: Section \ref{sec:system} introduces the communication system, explains various factors that affect the data rates of the system, and proposes an advanced modem to achieve higher data rates. Section \ref{sec:FIRsim} demonstrates the capabilities of the proposed system on FIR channel models in \cite{bos2018enabling}. Section \ref{sec:exp} expands the method to real world experiments. Finally, Section \ref{sec:conc} concludes the findings of the paper.



\section{Communication System}\label{sec:system}
In the proposed communication system (Fig. \ref{fig:framework}), the digital data stream is obtained as video data from a camera. The video bitstream obtained from the camera is mapped into transmit symbols by the digital modulator. The symbol sequence is mixed with the carrier and shifted to passband. The passband transmit signal is then realized by an arbitrary waveform generator and transmitted through the communication channel. In Section \ref{sec:FIRsim} the FIR channel response, and in Section \ref{sec:exp} the transmitting transducer, transmission medium (water or different biological tissues) and the receiving transducer constitute the communication channel. At the receiver end, the received waveform is sampled by a digitizer. Then, the received signal is moved to baseband and aligned coarsely before channel equalization is performed to compensate for the effects of the channel and to retrieve the transmitted symbols. Finally, the retrieved symbols are mapped back to the video data to be displayed on a screen.

\subsection{Communication Signal}
In a digital communications application, the channel is usually capable of transmitting a signal waveform instead of digital bits. If the channel is band-limited as in this case, the bandwidth of this transmission waveform should match, i.e. lie within, the channel's available bandwidth to not lose the transmitted information. The digital modulator maps the information bits to band-limited symbols that constitute the transmission waveform so that the waveform lies within the channel band.

The data rate in a digital communication system is given by
\begin{equation*}
    R = f_b \log_2 M
\end{equation*}
where $f_b$ is the symbol rate and $\log_2 M$ represents the number of bits in a symbol. $f_b$ is limited by the bandwidth of the communication channel. Hence, in order to achieve high data rates, the available bandwidth should be utilized as efficiently as possible with a high order modulation technique that can represent as many bits with one symbol as possible. 

In this paper, QAM is used due to its potential for high spectral efficiency. The modulator maps the binary data obtained from the webcam into $N$ symbols $\{x_1,\dots,x_N\}\in \{0,\dots,M-1\}$, which corresponds to $N\log_2 M$ bits. $M$ represents the order of QAM and indicates the number of possible complex symbol values the transmitted symbols may take. The symbols are upsampled by $L=\frac{f_s}{f_{b}}$, where $f_s$ is the sampling frequency of the digital-to-analog converter in the arbitrary signal generator, and shaped with a root-raised cosine filter $p(t)$, resulting in the data packet
\begin{equation}
    x_D(t)=\sum_{k=0}^{N-1}x_kp(t-kT_b)
\end{equation}
where $T_b=\tfrac{1}{f_b}$ is the symbol period.

In order to detect the signal arrival at the receiver, a preamble is appended at the beginning of each data packet at the transmitter, and the preamble is matched filtered at the receiver. This synchronizing preamble can be a chirp signal, or it can be a signal waveform representing a set of symbols. In this application, linear chirp is chosen as the preamble because it is more robust to Doppler \cite{stojanovic1994phase}. At the beginning of each data packet, linear chirp spanning from $-\tfrac{f_b}{2}$ to $\tfrac{f_b}{2}$ followed by a guard interval is appended. The guard interval is included to prevent the spreading of the preamble into the data packet due to long channel impulse response. The transmission packet, which consists of the linear chirp preamble, guard interval, and data packet, is then mixed with a sinusoidal carrier, where $f_c$ is the center frequency of the transmission band. The passband signal
\begin{equation}
    x(t) = \mathcal{R}e\left\{\sum_{k=0}^{N-1}x_kp(t-kT_b)e^{j2\pi f_c t}\right\}
\end{equation}
is then sent through the channel. 

\subsection{Communication Channel}
The communication channel consists of the transmitting transducer, transmission medium and the receiving transducer.
\subsubsection{Transducers}
The physical characteristics of transducers affect the data rates and the capability of the application in different ways. In a practical deployment scenario, the transducer is limited due to size constraints and the need to be biocompatible. Furthermore, the directivity of the source affects the SNR of the received signal, and the center frequency is related to the level of attenuation of the signal and the available bandwidth. 

\begin{itemize}
\item \textit{Directivity:} An essential property of a transducer is its ability to focus the transmitted energy in a particular direction and its sensitivity to the direction of the received signal. The more directional a transducer, the higher the projected and received signal power, resulting in higher SNR at the focal point. However, increased directivity requires the transmitting and receiving transducers' fields to be aligned precisely, or else the received signal might degrade significantly.
\item \textit{Center frequency:} The center frequency of a transducer, which dictates the frequency band of the passband signal, impacts the attenuation of the signal through a medium. In soft tissue at clinical ultrasonic frequencies, the attenuation increases with frequency. Therefore, higher frequencies give rise to lower signal power levels over the same transmission distance (Table \ref{tab:attenuation}). On the other hand, higher center frequency usually results in higher available bandwidth, which enables higher data rates. Therefore, it is crucial to choose an appropriate center frequency that balances the trade-off between attenuation, penetration depth and data rates.
\item \textit{Size:} The constraints imposed by the application typically limit the size of the transducer. For example, transducers that could be used in a wireless capsule endoscopy pill device would need to be smaller than the pill, whereas a permanently implanted device might have more flexibility in terms of size. 
\end{itemize}

\begin{table}[b]
\caption{Attenuation of ultrasonic waves at different frequencies traveling through different tissues of average thickness \cite{sayrafian2009statistical}}
\label{tab:attenuation}
\centering
\renewcommand{\arraystretch}{1.3}
\begin{tabular}{cl|p{.1\textwidth}|p{.1\textwidth}|}
\cline{3-4}
                                              &                 & \multicolumn{2}{c|}{\textbf{Attenuation (dB)}} \\ \cline{3-4}
\multicolumn{2}{l|}{}                                 & \hfil\textbf{1 MHz}             & \hfil\textbf{5 MHz}             \\ \hline
\multicolumn{1}{|c|}{\multirow{3}{*}{\rotatebox[origin=c]{90}{\textbf{Tissue}}}} & Fat             & \hfil1.52              & \hfil8.68              \\ \cline{2-4} 
\multicolumn{1}{|c|}{}                        & Muscle          & \hfil0.45              & \hfil4.24              \\ \cline{2-4} 
\multicolumn{1}{|c|}{}                        & Small Intestine & \hfil0.50               & \hfil2.49              \\ \hline
\end{tabular}%
\end{table}

\begin{figure}[t]
\includegraphics[width=.5\textwidth]{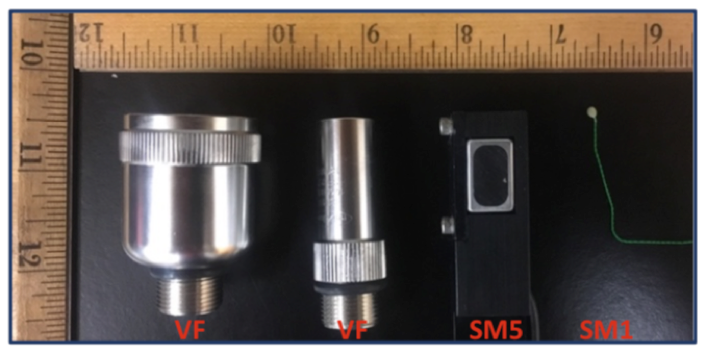}
\caption{Transmitting and receiving transducers with different center frequencies and form factors, with inch rulers for reference.}
\label{fig:transducers}
\end{figure}

\subsubsection{Transmission Medium}
The tissue type and the different segments that constitute the tissue affect the attenuation of the signal. In order to explore the effects of the transmission channel on the data rates, experiments were conducted through different media such as water, \textit{ex vivo} beef liver, \textit{ex vivo} pork chop, and \textit{in situ} through a rabbit abdominal wall. 
Beef liver and pork chop samples were initially used for this study because they are easy to obtain (e.g., can be bought from a grocery store). Furthermore, beef liver is a representative example of a homogeneous biological tissue while the pork chop is a non-homogeneous tissue consisting of layers with different attenuation and scattering properties.

Through water and the biological tissues used in this study, different frequencies travel with small variations in speed. More importantly, they are attenuated at different rates. In addition, the layers and the inhomogeneities within the tissues, as well as their reflective surroundings, cause multipath. As a result, the channel impulse response spans some number of symbols, causing intersymbol interference (ISI). Due to ISI, the transmitted symbols spread into each other at high data rates, and they are no longer distinguishable as separate pulses at the receiver.
\subsection{Channel Equalization}
\begin{figure}[t]
\includegraphics[width=\columnwidth]{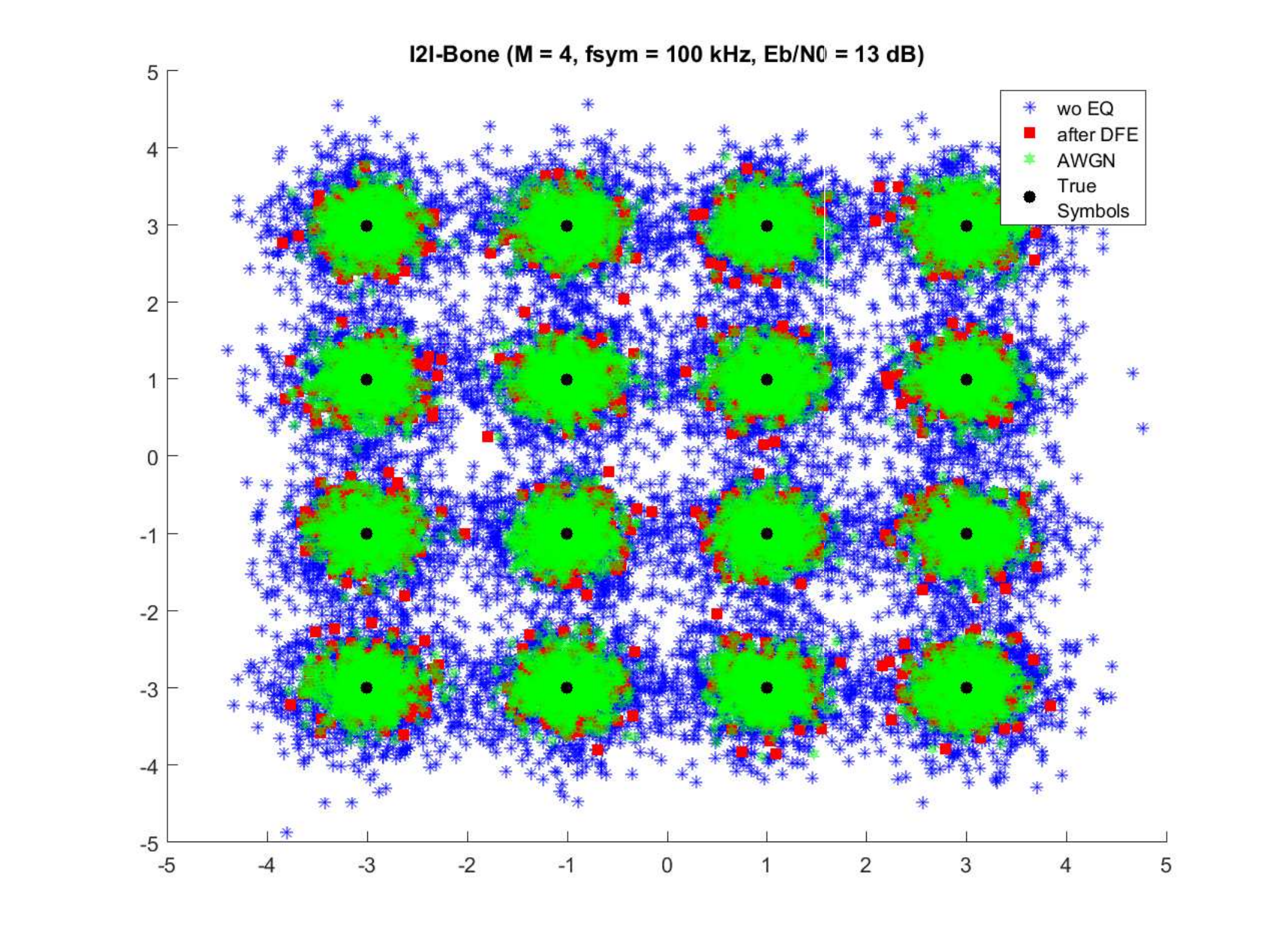}
\caption{Without equalization (blue stars), received symbols cannot be recovered successfully at higher data rates due to spreading effects of the channel on the transmitted symbols. The equalizer compensates for most of the channel effects and the received symbols after the equalizer (red squares) are better distinguishable. The symbols at output of the equalizer are comparable to the ones received through an AWGN channel (green diamonds).}
\label{fig:const_with_wothout_eq}
\end{figure}

The transducers and the transmission medium impose limits on the frequency components of the signal waveform that can pass through the channel. In order to achieve sufficient data rates for video streaming through bandlimited communication channels, higher-order modulation needs to be employed. However, because of dispersion, ISI, phase and frequency distortion, and noise, it is not possible to retrieve the higher-order transmitted QAM symbols at the receiver. Moreover, due to the long impulse response of the channel, without further processing, ISI becomes a further limiting factor in the data rates of the transmission system. Without equalization, received symbols cannot be demodulated successfully at higher data rates due to degrading effects of the channel on the transmitted signal (Fig. \ref{fig:const_with_wothout_eq}). The equalizer aims to compensate these effects of the channel in order to make the symbols separable and achieve better BER.

In this application, to account for the ISI while compensating for the channel in a computationally efficient way, a DFE \cite{austin1967decision} is used at the receiver end. The DFE is known to mitigate the effects of ISI by learning and feeding back an estimation of the ISI caused by the previous symbols on the current symbol estimated by a linear (feedforward) equalizer. 

A QAM signal requires coherent detection of the signal at the receiver for successful equalization and demodulation. Hence, a preamble is used to coarsely align the received signal. To obtain a reasonable coarse alignment that would fall within a few samples of the precise alignment, a preamble which has a high peak-to-sidelobe ratio in its autocorrelation function should be chosen. Furthermore, by choosing an appropriate preamble, a coarse Doppler estimate can be obtained, enabling the resampling of the signal at the receiver before demodulation and equalization. A linear chirp provides high peak-to-sidelobe ratio and it can also be used for coarse Doppler estimation \cite{stojanovic1994phase}. After coarse alignment and Doppler correction, the remaining Doppler effects and variations during the transmission can be compensated using a phase-locked loop in combination with a fractionally-spaced DFE \cite{stojanovic1994phase}. Once the transmitted symbols are estimated by the equalizer, they are mapped back to the bitstream and the corresponding video data.

\begin{figure*}[t]
     \centering
     \begin{subfigure}[t]{0.19\textwidth}
         \centering\includegraphics[width=\textwidth]{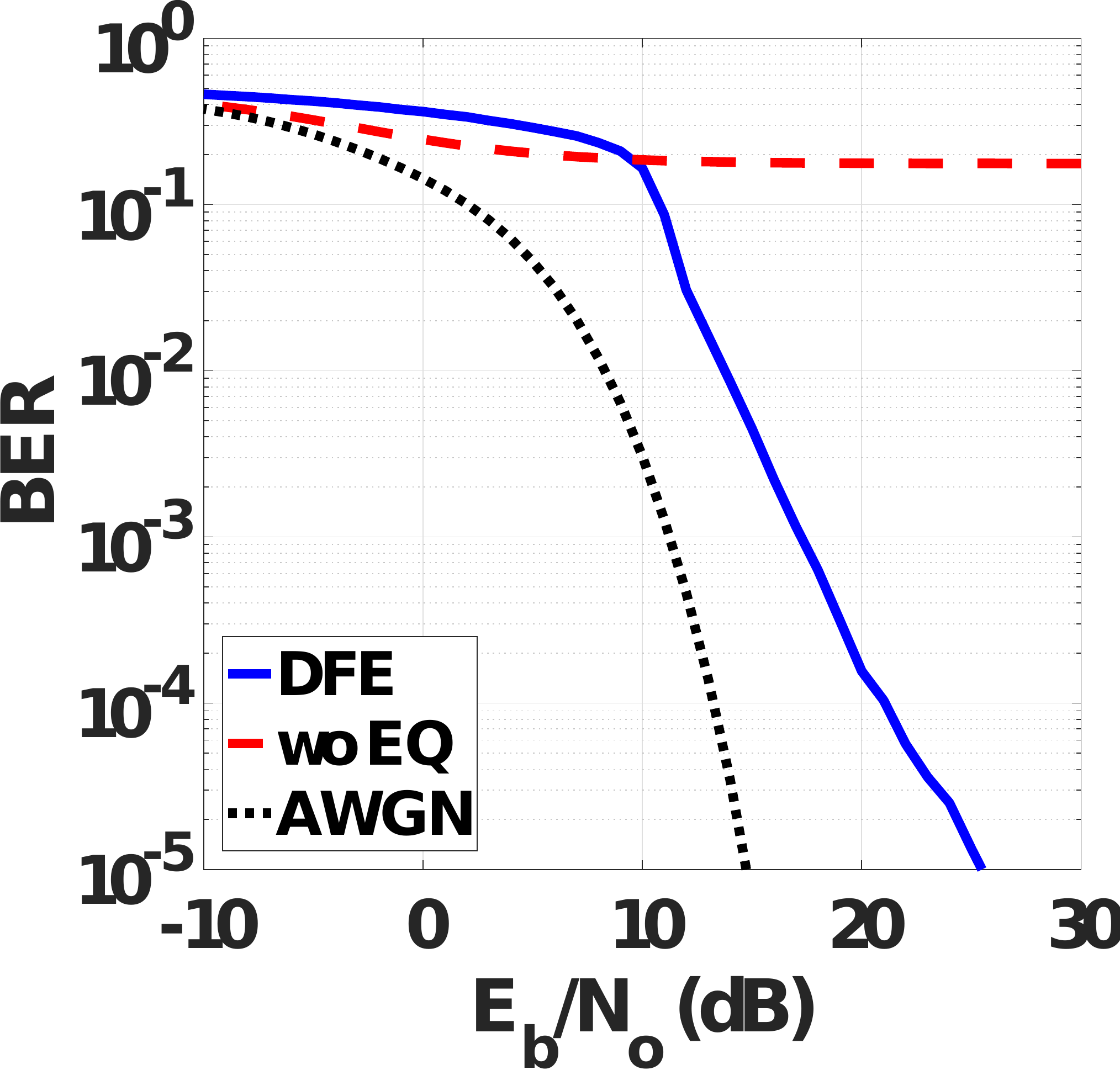}\label{fig:FIR_water_max}
         \caption[font=footnotesize]{Water (16-QAM, 500 kSymps)}
     \end{subfigure}
     \begin{subfigure}[t]{0.19\textwidth}
         \centering\includegraphics[width=\textwidth]{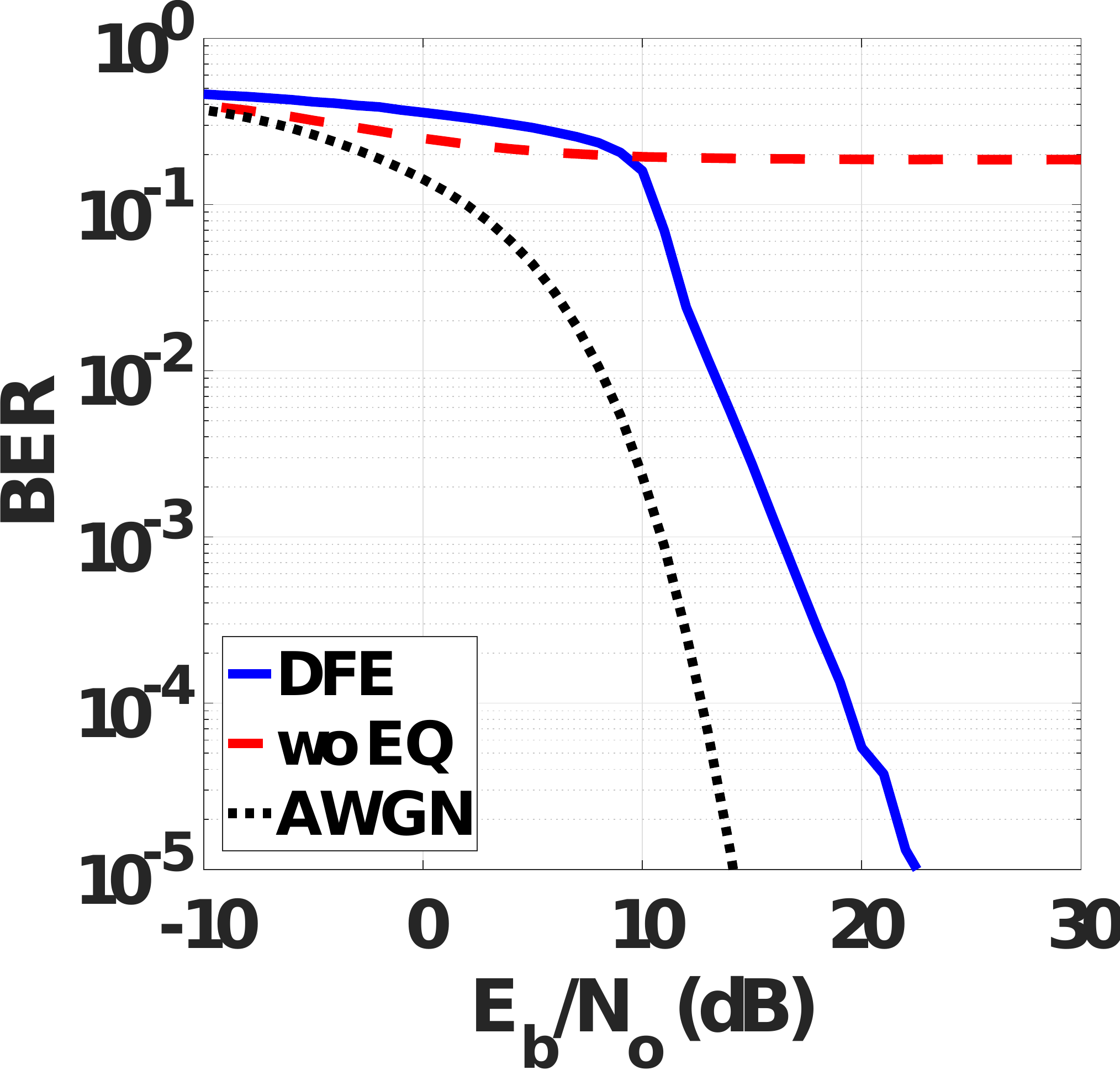}\label{fig:FIR_I2Ibone_max}
         \caption{I2I - Bone (16-QAM, 625 kSymps)}
     \end{subfigure}
     \begin{subfigure}[t]{0.19\textwidth}
         \centering\includegraphics[width=\textwidth]{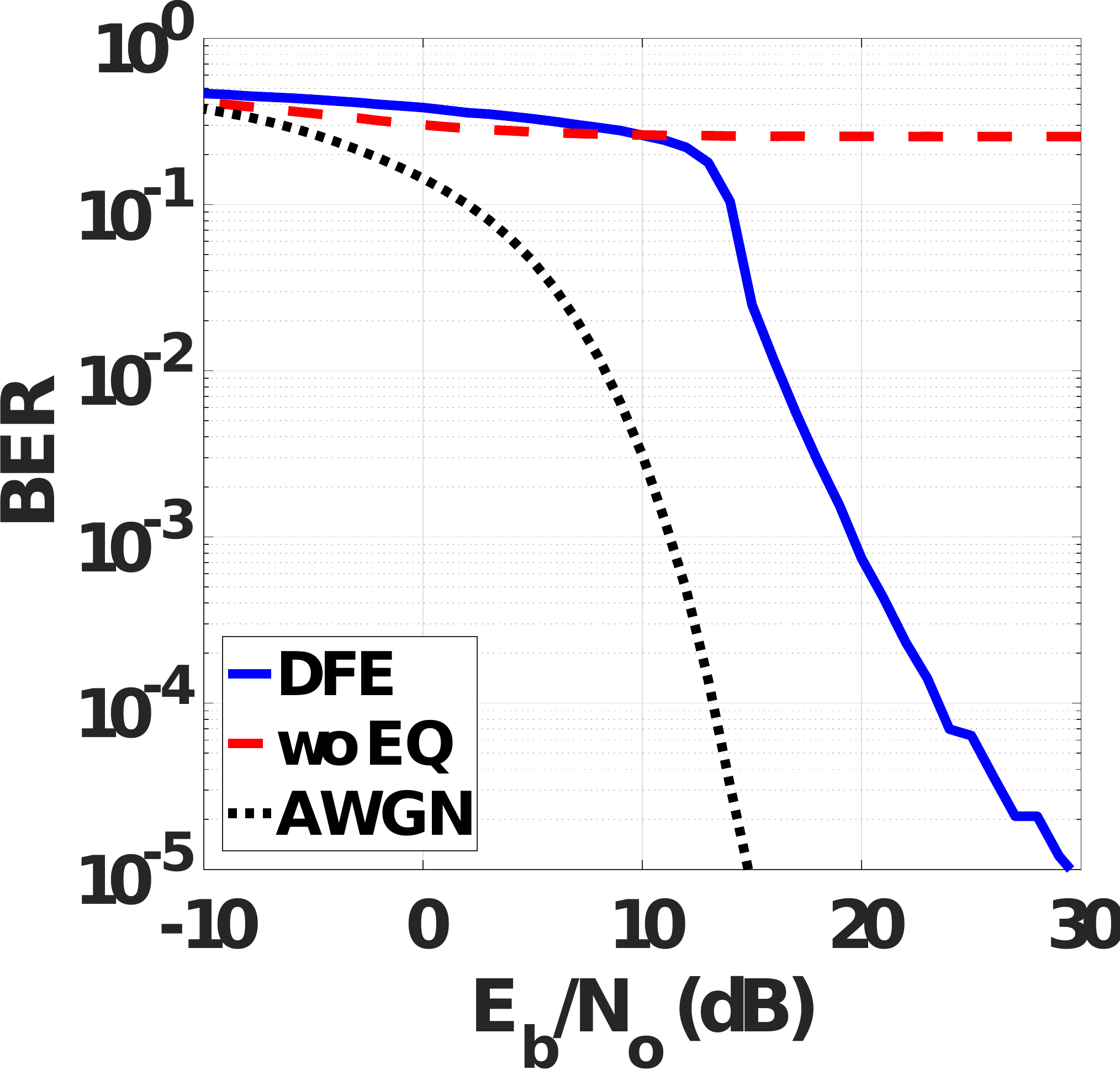}\label{fig:FIR_I2Igel_max}
         \caption{I2I - Gelatin (16-QAM, 500 kSymps)}
     \end{subfigure}
     \begin{subfigure}[t]{0.19\textwidth}
         \centering\includegraphics[width=\textwidth]{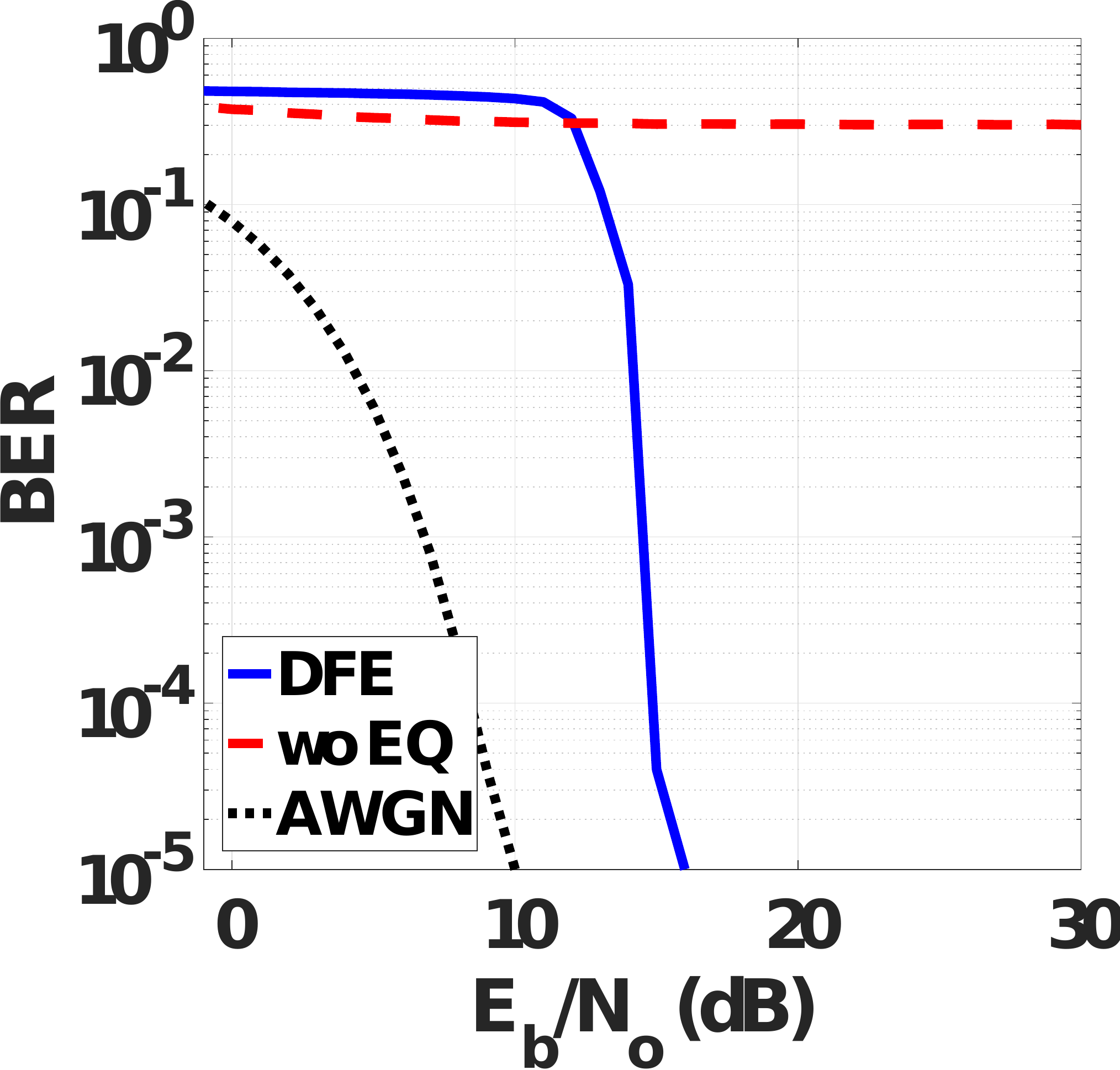}\label{fig:FIR_S2Sbone_max}
         \caption{S2S - Bone (QPSK, 100 kSymps)}
     \end{subfigure}
     \begin{subfigure}[t]{0.19\textwidth}
         \centering\includegraphics[width=\textwidth]{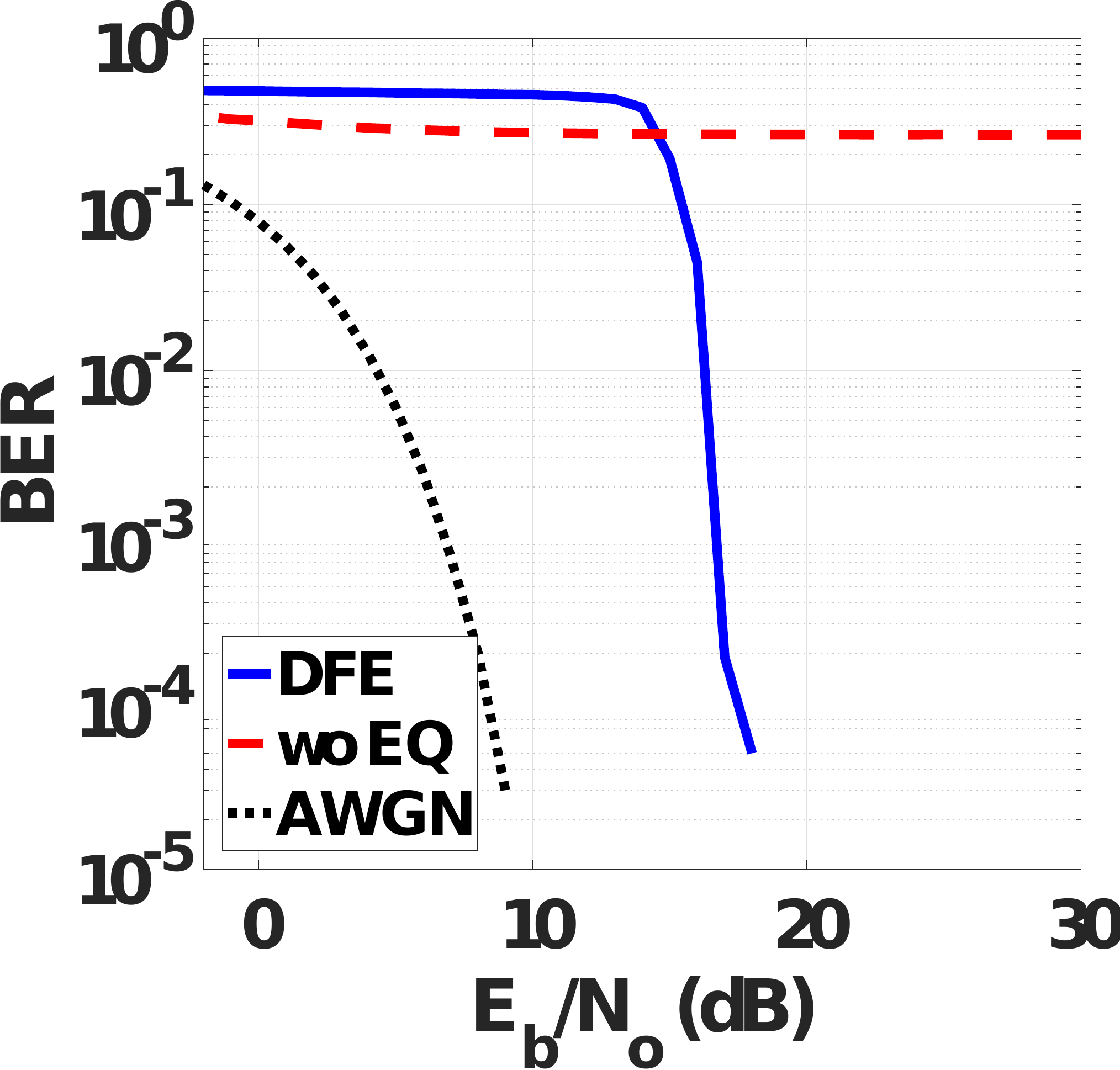}\label{fig:FIR_S2Sgel_max}
         \caption{S2S - Gelatin (QPSK, 100 kSymps)}
     \end{subfigure}
     \caption{BER versus $E_b/N_o$ plots of successful experiments that achieve the highest data rates using DFE for five different 80 mm channel models (solid blue), compared with decoding without equalization (dashed red) and AWGN channel (dotted black)}
     \label{fig:FIR_max}
\end{figure*}

\section{FIR Channel Simulations}\label{sec:FIRsim}
In order to determine the theoretical limits of the QAM modem and to compare the high data rate capabilities of the proposed method and the methods in the literature, a series of simulated experiments are performed with the FIR channel measurements provided by \cite{bos2018enabling}. The channel measurement dataset includes impulse response models of five different types of ultrasonic communication channels through (i) water, (ii) gelatin phantom with embedded transducers, mimicking implant to implant communication inside the body (I2I-Gelatin), (iii) gelatin-bone phantom with embedded transducers, mimicking implant to implant communication inside the body (I2I-Bone), (iv) gelatin phantom with transducers placed on the surface of the phantom, mimicking implant to implant communication on the surface of the body (S2S-Gelatin), and (v) gelatin-bone phantom with transducers placed on the surface of the phantom, mimicking implant to implant communication on the surface of the body (S2S-Bone). 
\subsection{Setup}\label{ssec:FIRsetup}
The communication system in Fig. \ref{fig:framework} is simulated by replacing the transducers and the propagation medium with the provided channel models for 80 mm propagation distance. The transducers used in \cite{bos2018enabling} are the same biocompatible 2-mm sonomicrometry crystals in this work. For each channel model, 16 experiments are repeated 100 times for each $M-$QAM, $M \in 2^{\{2, 4, 6, 8\}}$, and symbol rate $f_b \in \{100, 250, 500, 625\}$ kHz. The transmit data consisted of 50000 random bits. The bits mapped into corresponding QAM symbols and modulated on the carrier signal centered at $f_c=1.2$ MHz. The modulated waveform was preceded with a 10 microsecond linear chirp and one milisecond guard interval. The simulated received signals are obtained by adding white Gaussian noise with variance that corresponds to the desired SNR, measured per bit, represented as $E_b/N_o$. The received signal is first matched-filtered with the linear chirp to synchronize the signal arrival time, then fed into fractionally-spaced DFE, which had 18 $1/2$-spaced feedforward taps and 100 feedback taps. 10\% of the symbols are used for training, and DFE taps are updated with the recursive least-squares algorithm (RLS) with learning rate 0.997. 

\begin{table}[b]
\caption{Highest achieved data rates $R$ with BER\textless{}1e-4 for simulated experiments with 80 mm FIR channel models}
\label{tab:simrates}
\centering
\renewcommand{\arraystretch}{1.2}
\begin{tabular}{|l|x{.2\columnwidth}|x{.12\columnwidth}|x{.12\columnwidth}|x{.1\columnwidth}|}
\hline
\textbf{Channel}     &\textbf{Modulation} & \textbf{$\mathbf{f_b}$ (kHz)} & \textbf{$\mathbf{R}$ (kbps)} & \textbf{SNR (dB)} \\ \hline
\textbf{Water}       & 16-QAM                                   & 500                                             & 2000                                           &        22           \\ \hline
\textbf{I2I-Gelatin} & 16-QAM                                   & 500                                             & 2000                                           &        23           \\ \hline
\textbf{I2I-Bone}    & 16-QAM                                   & 625                                             & 2500                                           &        19           \\ \hline
\textbf{S2S-Gelatin} & QPSK                                     & 100                                             & 200                                            &        18           \\ \hline
\textbf{S2S-Bone}    & QPSK                                     & 100                                             & 200                                            &         15          \\ \hline
\end{tabular}
\end{table}

\begin{figure}[t]
     \centering
     \includegraphics[width=\columnwidth]{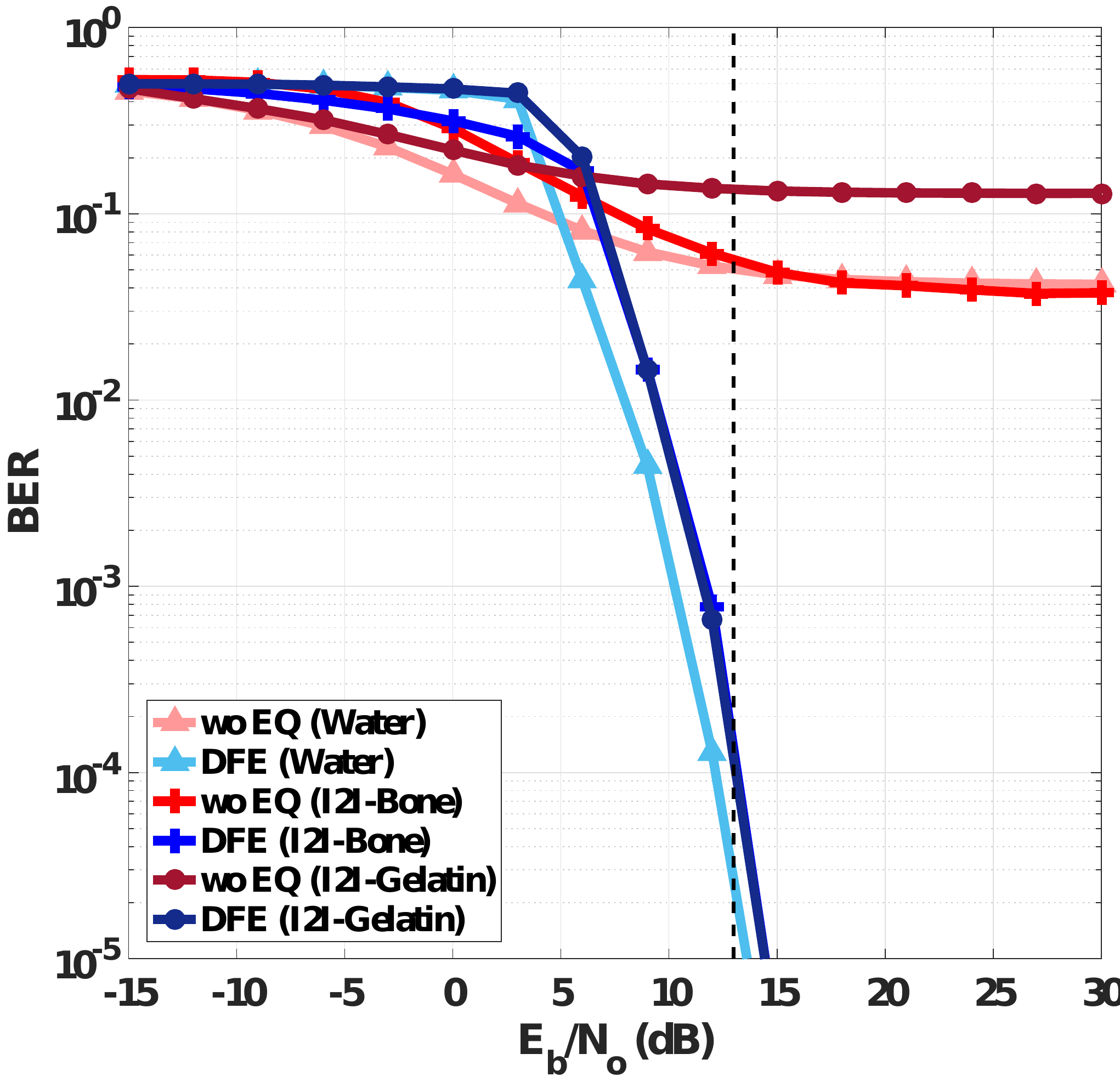}
     \caption{With the proposed DFE modem (DFE), 1 mbps is achieved with BER of 1e-4 at 13 dB. The same data rate results in BER$>$4e-2 with basic QAM modem (wo EQ) at 13 dB.}
     \label{fig:BERvsSNR_13db}
\end{figure}

\subsection{Results}
An experiment performed with an $(M, f_b)$ pair was considered successful if it could reach BER\textless{}1e-4 for $E_b/N_o$\textless{}30 dB. For each channel type, successful experiments that resulted in the highest data rates by utilizing the proposed method are summarized in Table \ref{tab:simrates}. For each successful experiment, BER versus $E_b/N_o$ curves in Fig. \ref{fig:FIR_max} were obtained by averaging the BER over 100 trials. For each trial, the received signal was also demodulated without equalization, which yields the results obtained with the basic QAM modem in \cite{bos2018enabling}. In order to establish the baseline for the AWGN channel, demodulation was performed on the noisy signal without channel effects. The proposed method achieved at least 2 mbps, providing 5 times increase in the data rates through water and 10 times increase through implant-to-implant communication channels compared to the basic QAM modem (QAM modulation and demodulation without equalization). Moreover, it enabled communication at 200 kbps through surface-to-surface channels, which was not possible with basic QAM modem. 

\begin{table*}[t]
\caption{Experimental transmission parameters, resulting data rates with BER\textless1e-4 and SNR at the output of the equalizer for transmissions through different media with different transducers}
\label{tab:results}
\centering
\renewcommand{\arraystretch}{1.3}
\begin{tabular}{|l|c|c|c|c|c|c|}
\hline
\textbf{Channel Type} & \textbf{Transducer} & \textbf{Modulation Format} & \textbf{$\mathbf{f_c}$ (MHz)} & \textbf{$\mathbf{f_b}$ (MHz)} & \textbf{Data Rate (Mbps)} & \textbf{SNR (dB)}                    \\ \hline
Water                 & VF1                 & 64-QAM                    &  1          &  1.25        &  7.5                      &  22 \\ \hline
Water                 & VF5                 & 64-QAM                     & 5           & 6.25           & 37.5                        & 20 \\ \hline
Water                 & SM1                 & 64-QAM                    & 1.25         & 1.1         & \textbf{6.7}              & 17                            \\ \hline
Water                 & SM5                 & 64-QAM                    & 5            & 5           & 30                        & 19                          \\ \hline
Beef Liver (2cm)      & VF1                 & 64-QAM                     & 1           &  1        &  6                      & 18 \\ \hline
Beef Liver (2cm)      & VF5                 & 16-QAM                     & 5           & 7.1         & 28.6                        & 15                          \\ \hline
Beef Liver (2cm)      & SM1                 & 16-QAM                    & 1.25         & 1.1         & \textbf{4.4}                & 13 \\ \hline
Beef Liver (2cm)      & SM5                 & 64-QAM                     & 5           & 4.2         & 25                        & 18 \\ \hline
Beef Liver (5cm)      & VF1                 & 64-QAM                     & 1           &  1        &  6                      & 17 \\ \hline
Beef Liver (5cm)      & VF5                 & 16-QAM                     & 5           & 5.5         & 22.2                        & 16 \\ \hline
Beef Liver (5cm)      & SM1                 & 16-QAM                    & 1.25         & 1.1         & \textbf{4.4}                & 14                           \\ \hline
Beef Liver (5cm)      & SM5                 & QPSK                     & 5           & 5         & 10                        & 11                          \\ \hline

\end{tabular}%
\end{table*}

The required SNR for 200 kbps through implant-to-implant channels with a basic QAM modem is reported as 13 dB in \cite{bos2018enabling}. Although SNR values required to achieve the data rates in Table \ref{tab:simrates} are higher than 13 dB, the proposed method achieved 1.25 mbps and 1 mbps through water and implant-to-implant channels at 13 dB (Fig. \ref{fig:BERvsSNR_13db}), providing 3 to 5 times improvement compared to the basic QAM modem.

In accordance with the findings in the literature, the simulations demonstrated that the basic QAM modem is not capable of communicating at high data rates required for video transmission through biological tissues with small, biocompatible transducers for given channel models. An advanced equalizer tailored for this particular application, on the other hand, shows promise towards the video-capable data rates through biological tissues at moderate to high SNR levels. 

\section{Experiments}\label{sec:exp}
In order to test the capabilities of the proposed method on real data, two sets of experiments were performed on \textit{ex vivo} and \textit{in situ} biological tissues. In the first set of experiments, different transducers were used to communicate through water and different thicknesses of \textit{ex vivo} beef liver to explore the capabilities of the method with transducers that have different bandwidths and received signal powers through similar type of transmission media. In the second set of experiments, small, biocompatible transducers were used through \textit{ex vivo} pork chop and \textit{in situ} through the rabbit abdomen to explore the capabilities of the method through different media and surrounding environments.

\subsection{Different transducers, similar channels}\label{ssec:waterbeef}
\subsubsection{Setup}
\begin{figure}
     \centering
     \begin{subfigure}[t]{0.48\linewidth}
         \centering
         \includegraphics[width=\textwidth]{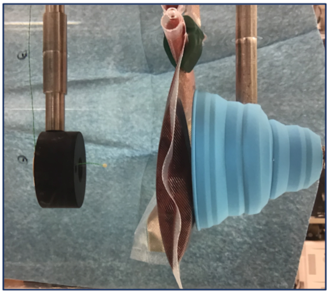}
         \caption{2mm @ 1.3 MHz transducer (SM1) through 2 cm beef liver}
         \label{fig:BL_SMC}
     \end{subfigure}
     \hfill
     \begin{subfigure}[t]{0.48\linewidth}
         \centering
         \includegraphics[width=\textwidth]{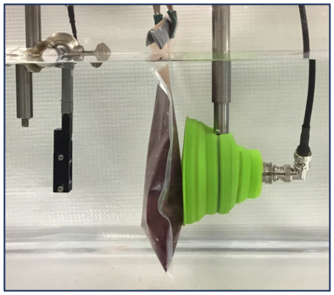}
         \caption{5$\times$8 mm$^2$ @ 5 MHz transducer (SM5) through 2 cm beef liver}
         \label{fig:BL_S5M}
     \end{subfigure}
     \caption{Experimental setup with different transducers suspended in a water tank, facing each other with beef liver in between. A funnel is used to prevent potential paths reflecting from the tank's walls and arriving at the receiver without passing through the tissue.}
     \label{fig:BL_setup}
\end{figure}

The first set of experiments were performed in a water tank filled with degassed water. In the water tank, two transducers were placed in a pitch-catch configuration. In the through-water experiments, the plastic bag, in which the beef liver was placed in further experiments, was filled with degassed water and suspended between the transducers to account for the reflective properties of the bag. In the experiments with beef liver, \textit{ex vivo} beef livers of different thicknesses placed in between the transducers (Fig. \ref{fig:BL_setup}). The experiments were conducted using four different transducers (Fig. \ref{fig:transducers}) that have different sizes, directivities, and center frequencies in order to examine the effects of each of these factors on data rates. The transducers used in the experiments were: large, focused and directional transducers with 1 and 5 MHz center frequencies (Valpey Fisher, Hopkinton, MA), denoted as \textit{VF1} and \textit{VF5}, biocompatible sonomicrometry crystals of 2-mm diameter that operate around 1.3 MHz (Sonometrics, London, CAN), denoted as \textit{SM1}, and a rectangular 5 MHz transducer with 5 mm width and 8 mm height denoted as \textit{SM5}. At the receiver end, because the applications are typically not limited by the transducer size outside of the body, large and highly focused transducers with 1 and 5 MHz center frequencies, \textit{VF1} and \textit{VF5}, were used to maximize the received signal power. The receiving transducer was also placed inside a soft, plastic funnel to verify that the signal does not arrive at the receiving transducer via potential signal paths reflected from the tank's walls without traveling through the biological tissue.

The transmission data were generated from random bits with different symbol rates and modulation orders as described in Section \ref{ssec:FIRsetup}. For each experiment setup, two packets of 20,000 symbols each were generated randomly with chirp and guard interval preceding each packet. In all the experiments, the signal was generated, transmitted, recorded, and then processed. An arbitrary waveform generator (PXI-5422, National Instruments, Austin, TX) was used to generate the transmission signal at the preset center frequencies with different bandwidths, and coded with different modulation schemes. A digitizer was used (PXI-5124, National Instruments, Austin, TX) to acquire the signal at the receiver end. In order to drive the National Instruments equipment and to process the received data, custom Matlab (MathWorks, Natick, MA) software was used. 

At the receiver end, the signal was captured by the digitizer, coarsely aligned and corrected for Doppler effects using the chirp preamble, and the received data packet was decoded using the fractionally-spaced, phase-tracking, sparse DFE \cite{stojanovic1994phase}, \cite{lopez2001dfe}. The equalizer had at most 53 feedforward taps and 90 feedback taps. Out of 20,000 symbols transmitted in each packet, 10\% were used for training (except when the channel response was longer through water with \textit{VF5}, which required 14\% training) to learn the equalizer coefficients, and the rest were used in the decision-directed mode while updating the equalizer coefficients to track the channel variations. The coefficients were updated using the recursive least-squares algorithm with learning rate 0.997.

\begin{figure}[b]
     \centering
         \includegraphics[width=.5\textwidth]{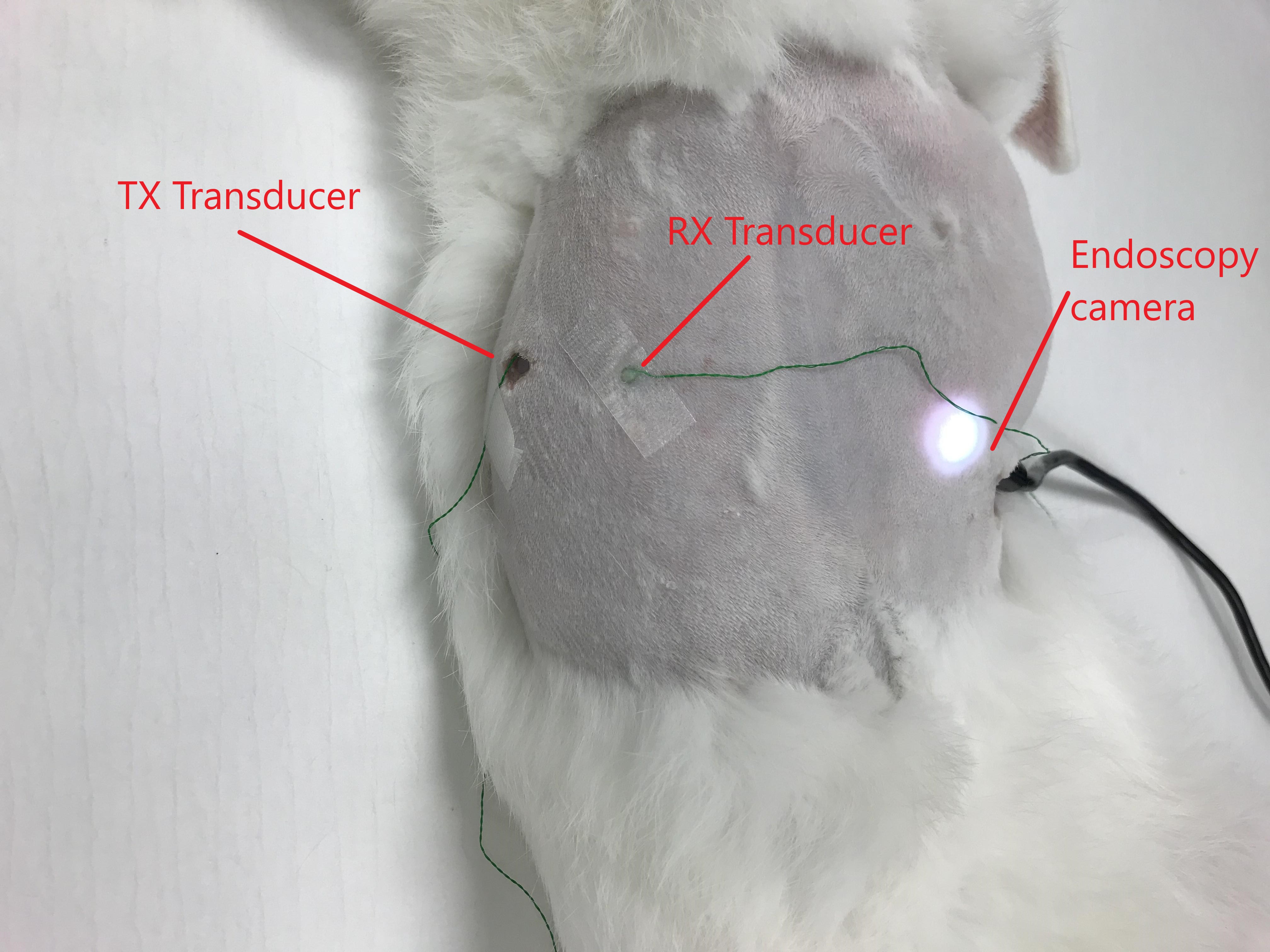}
     \caption{Transmission of the data, obtained with an endoscopy camera, through the abdominal wall of a rabbit using 2 mm sonomicrometry crystals}
     \label{fig:rabbit}
\end{figure}

\subsubsection{Results}
The results of the first set of transmission experiments through water and beef liver are listed in Table \ref{tab:results}. The thickness of the biological tissue increases attenuation, and results in decreased data rates for a given transmission frequency. For a given thickness, higher frequencies provide higher data rates despite the increased attenuation, thanks to higher available bandwidth. However, the form factors of the high-frequency transducers are too large to be used in an IMD such as video capsule endoscopy pill. Hence, it is more realistic to focus on the small form factor transducers (\textit{SM1}), for which the experiments demonstrate video-capable data rates through water and beef liver.

\begin{figure}[t]
     \centering
     \begin{subfigure}[t]{0.49\linewidth}
         \centering\includegraphics[width=\textwidth]{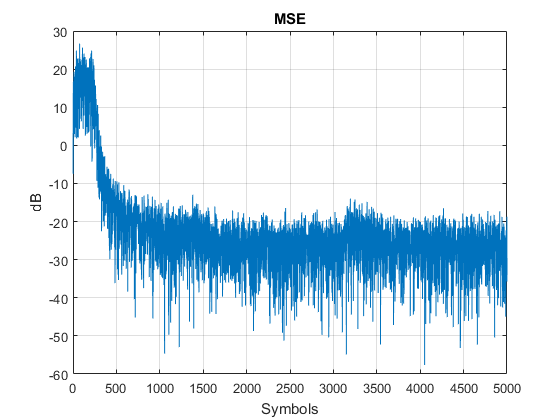}
     \end{subfigure}
     \begin{subfigure}[t]{0.49\linewidth}
         \centering\includegraphics[width=\textwidth]{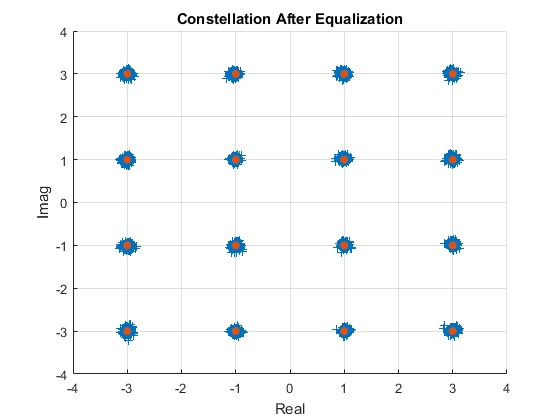}
     \end{subfigure}
     \begin{subfigure}[t]{0.49\linewidth}
         \centering\includegraphics[width=\textwidth]{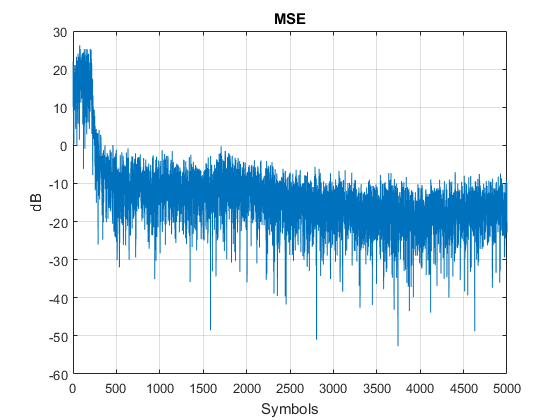}
     \end{subfigure}
     \begin{subfigure}[t]{0.49\linewidth}
         \centering\includegraphics[width=\textwidth]{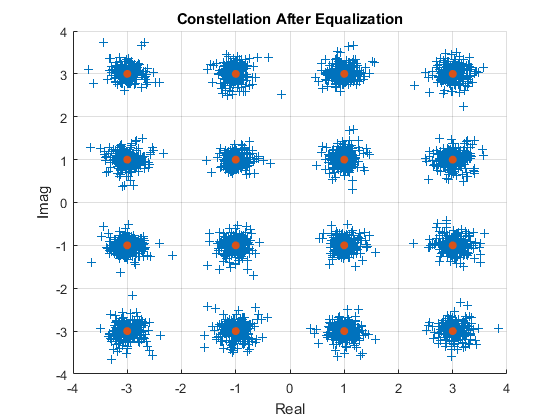}
     \end{subfigure}
     \begin{subfigure}[t]{0.49\linewidth}
         \centering
         \includegraphics[width=\textwidth]{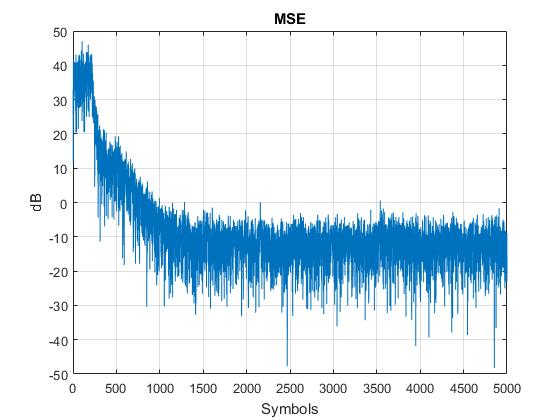}
     \end{subfigure}
     \hfill
     \begin{subfigure}[t]{0.49\linewidth}
         \centering
         \includegraphics[width=\textwidth]{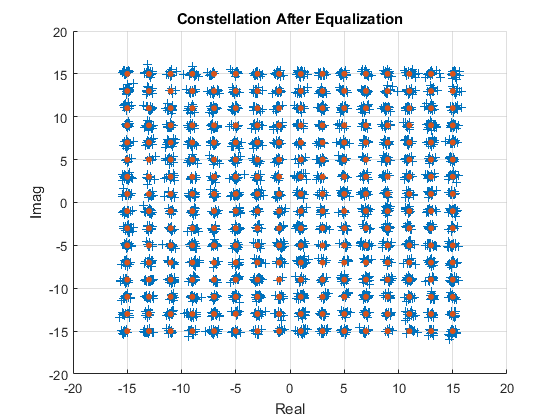}
     \end{subfigure}
     \caption{Mean-squared error (left) and received signal constellation after equalization (right) for transmissions through 2 cm \textit{ex vivo} pork chop (top), 8 cm \textit{ex vivo} pork chop (middle) and \textit{in situ} rabbit abdomen (bottom)}
     \label{fig:constellation}
\end{figure}

\subsection{Different channels, 2-mm transducers}
\subsubsection{Setup}
In order to examine the effects of tissue properties on the data rates, the second set of experiments were performed with 2-mm transducers (\textit{SM1}) communicating through \textit{ex vivo} pork chop and \textit{in situ} through the rabbit abdomen. First, two \textit{SM1} transducers were attached at each sides of 2 cm and 8 cm pork chops suspended in air. The transmission data was obtained from a webcam (c925e USB HD Webcam, Logitech, Lausanne, CH) that has onboard h.264 video compression. Through 2 cm and 8 cm pork chops, 16-QAM, 1.4 MHz center frequency signal with 500 kHz symbol rate and 16-QAM, 1.4 MHz center frequency signal with 400 kHz symbol rate were transmitted, respectively.

The experiments were repeated by streaming video with SM1 transducers through the abdominal wall of a euthanized rabbit to demonstrate \textit{in situ} transmission capabilities of the system. The transmitting transducer was implanted behind the abdominal wall, and the receiving transducer was placed on the shaved abdomen of the rabbit with gel coupling (Fig. \ref{fig:rabbit}). The data was obtained through an endoscopy camera (T01 8.5 mm USB Semi-Rigid Endoscope, Depstech) placed inside the abdomen of the rabbit. The data was sent through the rabbit abdomen with a 256-QAM, 1.1 MHz center frequency signal with 400 kHz symbol rate. A similar receiver structure to Section \ref{ssec:waterbeef} was used for each experiment.

\subsubsection{Results}
The mean-squared error plots and constellation diagrams for received, equalized symbols sent through the pork chops are displayed in the top and middle plots of Fig. \ref{fig:constellation}. The results demonstrate successful demodulation of the symbols for each experiment. Although the current experimental results achieved video-capable data rates of 2 mbps and 1.6 mbps with BER$<$6.25e-5, the $E_b/N_o$ at the output of the equalizers in these experiments suggest that successful demodulation would be possible with even higher order of QAM. Specifically, 28 dB output $E_b/N_o$ for the 2 cm pork chop experiment would allow 256-QAM, yielding data rates of at least 4 mbps, and the 16.4 dB output $E_b/N_o$ for the 8 cm pork chop would allow 64-QAM, yielding 2.4 mbps \cite{proakis2001digital}. 
Finally, the bottom plots in Fig. \ref{fig:constellation} display the successful transmission through the rabbit abdomen, which results in data rates of 3.2 mbps with BER$<$4.2e-5.

The \textit{ex vivo} pork chop and \textit{in situ} rabbit experiments demonstrated that it is possible to achieve video-capable data rates through different biological tissues using small, biocompatible transducers that could be utilized in a packaged IMD.

\section{Conclusion}\label{sec:conc}
An online, wireless IMD that is capable of video communications or with the ability to transfer data at rates necessary for video streaming would be revolutionary in medical therapy and diagnostics. These demonstrated rates would allow the IMDs to live-stream images or download software updates, patient history, or other collected data, all within a single office visit. The regulations on and characteristics of RF electromagnetic waves restricts their use in wireless IMDs. 

Previous works in the literature demonstrated the feasibility of ultrasonic waves to communicate through the body. However, the realization of such communications at high data rates with small form factor transducers were pending. In this work, it is demonstrated for the first time that achieving video-capable data rates using ultrasonic waves through different biological tissues is possible by employing refined signal processing and communication techniques.

\begin{table}[t]
\caption{Data rates with corresponding bit error rates through different tissues using 2 mm SM1 transducers}
\begin{minipage}{.5\textwidth}
\centering
\renewcommand{\arraystretch}{1.3}
\begin{tabular}{|l|c|c|}
\hline
\textbf{Channel Type} & \textbf{Data Rate} & \textbf{BER}      \\ \hline
Water                 & 6.7 mbps           & 5e-5             \\ \hline
Beef Liver (2 cm)      & 4.4 mbps             & 3e-5   \\ \hline
Beef Liver (5 cm)      & 4.4 mbps             & 4e-5              \\ \hline
Pork Chop (2 cm)       & 4 mbps\footnote{\label{note1}{Extrapolating from the output SNR}}             & \textless{}1e-6 \\ \hline
Pork Chop (8 cm)       & 2.4 mbps\textsuperscript{\ref{note1}}         & \textless{}1e-3 \\ \hline
Rabbit Abdomen        & 3.2 mbps           & \textless{}4.2e-5 \\ \hline
\end{tabular}
\vspace{-0.2cm}
\end{minipage}
\label{tab:data_rates}
\end{table}

\section*{Acknowledgment}

This work was supported by a grant from the National Institutes of Health (NIH R21EB025327).

\ifCLASSOPTIONcaptionsoff
  \newpage
\fi



\bibliographystyle{IEEEtran}
\bibliography{meatcomms2_preprint}
\end{document}